\documentclass[aip,apl,reprint]{revtex4-1}
\usepackage{graphicx}
\usepackage{bm}
\usepackage{textcomp}
\usepackage{color}

\begin{document}

\title{Frequency Division Multiplexing Readout and Simultaneous Manipulation of an Array of Flux Qubits}

\author{M.~Jerger}
\affiliation{Physikalisches Institut, Karlsruhe Institute of Technology and DFG-Center for Functional Nanostructures (CFN),
D-76128 Karlsruhe, Germany}
\author{S.~Poletto}
\altaffiliation[Present address: ]{IBM T.J. Watson Research Center, PO Box 218,
Yorktown Heights, NY 10598, USA}
\affiliation{Physikalisches Institut, Karlsruhe Institute of Technology and DFG-Center for Functional Nanostructures (CFN),
D-76128 Karlsruhe, Germany}

\author{P.~Macha}
\affiliation{Institute of Photonic Technology, PO Box 100239, D-07702 Jena, Germany}

\author{U.~H\"ubner}
\affiliation{Institute of Photonic Technology, PO Box 100239, D-07702 Jena, Germany}

\author{E.~Il'ichev}
\affiliation{Institute of Photonic Technology, PO Box 100239, D-07702 Jena, Germany}

\author{A.~V.~Ustinov}
\email{ustinov@kit.edu}
\affiliation{Physikalisches Institut, Karlsruhe Institute of Technology and DFG-Center for Functional Nanostructures (CFN),
D-76128 Karlsruhe, Germany}

\date{\today}

\begin{abstract}
An important desired ingredient of superconducting quantum circuits is a readout scheme whose complexity does not increase with the number of qubits involved in the measurement. Here, we present a readout scheme employing a single microwave line, which enables simultaneous readout of multiple qubits. Consequently, scaling up superconducting qubit circuits is no longer limited by the readout apparatus. Parallel readout of 6 flux qubits using a frequency division multiplexing technique is demonstrated, as well as simultaneous manipulation and time resolved measurement of 3 qubits. We discuss how this technique can be scaled up to read out hundreds of qubits on a chip.
\end{abstract}



\maketitle

Two key requirements for realizing a quantum computer are the abilities to manipulate a register of qubits and to measure its state. 
Nowadays, registers of up to 3-4 superconducting qubits have been manipulated and read out simultaneously either by using a common resonator\cite{DiCarlo-Nat-10} or by employing dedicated electronics for each qubit\cite{Martinis-Nat-10}.
Both techniques are difficult to scale up towards quantum processors with a large number of qubits which is required for implementing practically relevant computational tasks. 

In this letter, we describe and demonstrate an approach that combines parallel manipulation and readout of multiple qubits using a frequency division multiplexing (FDM) scheme. FDM has previously been applied in other fields, e.g. for reading out arrays of microwave kinetic inductance detectors\cite{Mazin-NIM-06,Yates-APL-09}. 
The advantage of this technique is that it is scalable, in principle, to an arbitrarily large number of devices. 
The architecture employs resonators of different frequencies for individual qubits\cite{Jerger-EPL-2011}, all addressed using a single common microwave transmission line that connects the qubit chip to the readout electronics. 

The frequency-selective approach\cite{Jerger-EPL-2011} to read out $N$ qubits would require $N$ individual microwave frequencies for readout, which also suffers from scaleability issues if generated using individual microwave generators. 
This engineering problem can be solved by using techniques from software-defined radio (SDR), a core technology of many modern wireless transceivers.
On the readout side, a fast digital-to-analog converter (DAC) synthesizes $N$ specific sub-GHz baseband signal tones for individual qubits, as many tones as desired within the available bandwidth. 
Subsequently, these tones are mixed up to the desired readout resonator frequency of few GHz by using a reference microwave source and a two-quadrature (IQ) mixer. 
The generated multi-tone probe signal is sent through an on-chip transmission line with individual qubit readout resonators coupled to it. 
The transmitted signal is mixed back down to baseband frequencies and digitized by a fast analog-to-digital converter (ADC), and the amplitude and phase change of each transmitted tone are obtained from the quadratures I and Q using a fast Fourier transform.
Quantum manipulation and readout of $N$ qubits requires $N$ microwave tones to be generated in addition to the readout tones. The manipulation tones can be generated using the same SDR technique under the condition of all qubit manipulation frequencies remaining in a range of two times the available baseband bandwidth around a common reference frequency. 

The investigated circuit consists of seven flux qubits\cite{Orlando-PRB-99}, each inductively coupled to an individual coplanar waveguide resonator. 
The seven resonators were designed to have distinct resonance frequencies ranging from 9.3 GHz to 10.3 GHz. 
Each resonator is capacitively coupled to a common $50\,\rm \Omega$ coplanar transmission line crossing the whole chip. 
Details on the sample layout and fabrication can be found elsewhere\cite{Jerger-EPL-2011}. 
Our test chip was designed to demonstrate FDM; the qubits are not directly coupled to each other. 
The circuit had no on-chip magnetic coils to individually bias each qubit of the array. 
The qubits were biased to their operating points via a uniform field coil wrapped around the sample holder and two small gradient coils placed above the sample. 
The three coils allow simultaneous operation of three qubits at their symmetry points. 
Qubit manipulation was performed by an additional external microwave signal applied via the readout transmission line.

\begin{figure*}[t]
\begin{center}
\includegraphics{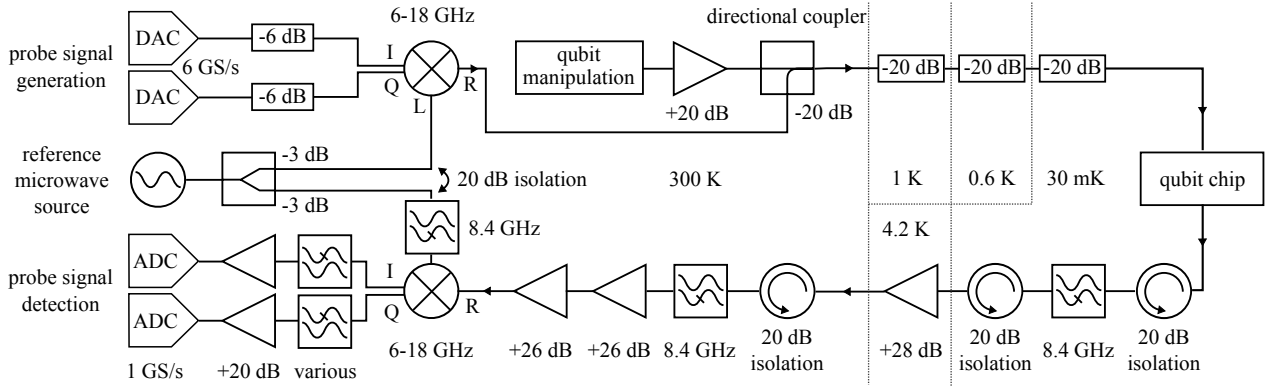}
\end{center}
\caption{Experimental setup used for FDM readout. The qubit manipulation signal is generated by a single microwave source for spectroscopy and three additional microwave sources, DAC channels and mixers for pulsed excitation of the qubits.}
\label{fig:setup}
\end{figure*}

A probe signal composed of one microwave tone per qubit to be read out is sent through the common transmission line. 
The interaction between each qubit and resonator leads to a state-dependent dispersive shift\cite{Blais-PRB-04}, $\Delta\omega_r = \tilde{g}^2 / (\omega_q-\omega_r) \sigma_z$, of the resonator frequency, where $\tilde{g}$ is the effective coupling between the resonator and qubit, $\omega_q$ and $\omega_r$ are the angular resonance frequencies of the qubit and resonator, and $\sigma_z$ is $\pm 1$ depending on the state of the qubit.
The composite signal probes all resonators at the same time, storing the information on the state of all qubits in the transmitted tones. 
Detection of the transmitted amplitude and phase of each of the tones provides a simultaneous non-destructive measurement of the states of all qubits. 
The probe signal is generated by mixing a reference microwave tone in the band of the resonators and a multi-tone DAC output using an IQ mixer, see Fig.~\ref{fig:setup}. 
By using the I and Q quadratures, we address the upper and lower sidebands of the mixing product individually to effectively double the bandwidth of the system. 
The mixer output is combined with the qubit manipulation signal through a directional coupler.
A strongly attenuated line transmits the combined signal to the sample, which is attached to the mixing chamber stage of a dilution refrigerator.
Two cryogenic circulators and a high-pass filter at 30\,mK are used to prevent reflections and noise from traveling from the cryogenic amplifier back to the sample. 
A chain of amplifiers provides 80\,dB gain to boost the transmitted probe signal to levels sufficient for the detection stage, which employs an identical IQ mixer to convert the signal back to baseband frequencies. 
The local oscillator inputs of both mixers are fed from the same reference microwave source, resulting in a homodyne detection with a fixed phase offset. 
An additional high-pass filter between the local oscillator ports of the two mixers prevents leakage of the baseband signal.
After digitizing both quadratures, the amplitude and phase of all components of the probe signal are extracted via FFT. 
The maximum number of devices that can be probed with the described technique is defined by the frequency separation between resonators and the bandwidth of the acquisition board. 
The frequencies of the resonators on our chip are spaced at intervals of 150\,MHz and the acquisition board has a bandwidth somewhat below 500\,MHz, allowing for the simultaneous detection of up to six devices. 

\begin{figure}[b]
\begin{center}
\includegraphics{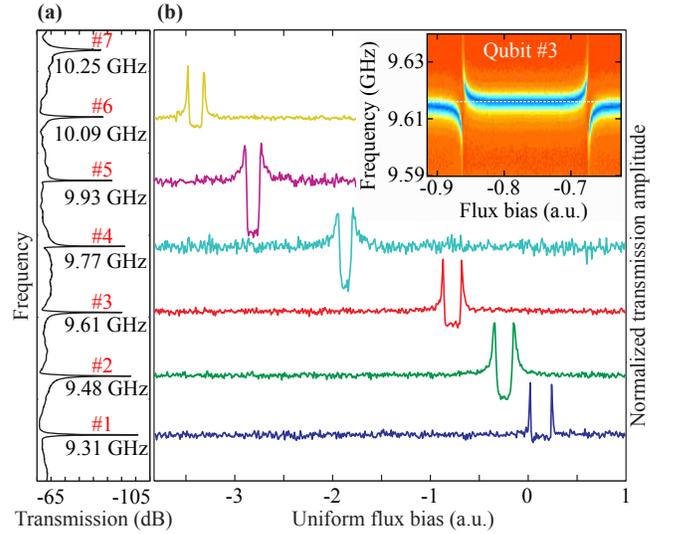}
\end{center}
\caption{
  (color online). 
  (a) Transmission spectrum of the sample with all qubits far detuned from the resonances. 
  (b) FDM readout of six flux qubits. 
    The main plot shows the transmission amplitude at the resonance frequencies of devices \#1 to 6 vs. the magnetic flux generated by the uniform field coil, measured using FDM.
    The inset shows the transmission amplitude at several frequencies close to resonance \#3, measured with a network analyzer. 
The dashed line indicates the probe frequency used for this device in the main plot.
    The curves in the main plot are normalized and shifted vertically for better visibility. 
    The offset along the horizontal axis is due to magnetic field non-uniformity, which is likely due to the screening currents generated in the superconducting ground plane. 
}
\label{fig:tm6q} 
\end{figure}

The transmission spectrum of the sample, measured with a vector network analyzer is reported in Fig.~\ref{fig:tm6q}(a). 
The seven absorption peaks correspond to the seven readout resonators. 
Close to each peak its bare resonance frequency as well the identification number of the device are printed. 
The inset of Fig.~\ref{fig:tm6q}(b) shows the transmission at frequencies around the resonance of device \#3 vs. the magnetic flux bias.
The two points at which the dispersive frequency shift changes its sign correspond to avoided level crossings of the qubit and resonator.
We demonstrate FDM by measuring the maximum possible number of devices simultaneously. 
Figure~\ref{fig:tm6q}(b) shows the transmitted amplitude of the six probe tones versus the external uniformly applied magnetic flux. 
Each curve is shown aligned with the corresponding transmission peak to the left in Fig.~\ref{fig:tm6q}(a). 
The amplitude of the transmitted signal is constant as long as the qubit remains far detuned from the resonator. 
The amplitude changes drastically around two distinct fluxes, again indicating anti-crossings between the qubit and the corresponding resonator. 
There is a minimum between these two peaks, because the readout frequencies were set on resonance, with the dispersive shifts at the symmetry points of the qubits taken into account.
The readout frequency of device \#3 is shown as a dashed line in the inset.

\begin{figure}[tb]
\begin{center}
\includegraphics{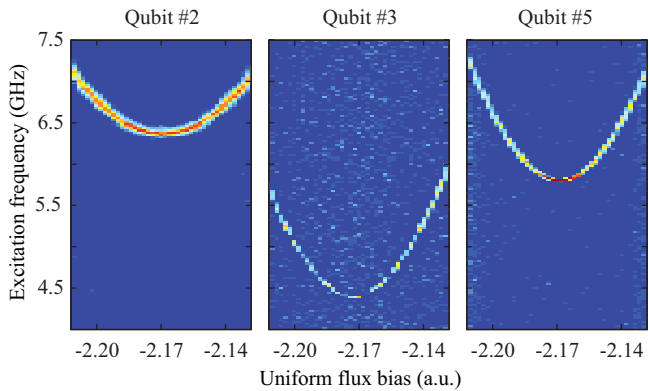}
\end{center}
\caption{
  (color online). 
  Multiplexed spectroscopy of qubits \#2, 3 and 5. 
  The qubit manipulation microwave excites qubits when its frequency matches the transition between their ground and excited states. 
  The state of all three qubits is continuously and simultaneously monitored by the multi-tone probe signal. 
  The horizontal axis reports the uniform bias flux applied to the chip.
}
\label{fig:multi_spec}
\end{figure}

\begin{figure}[t]
\begin{center}
\includegraphics{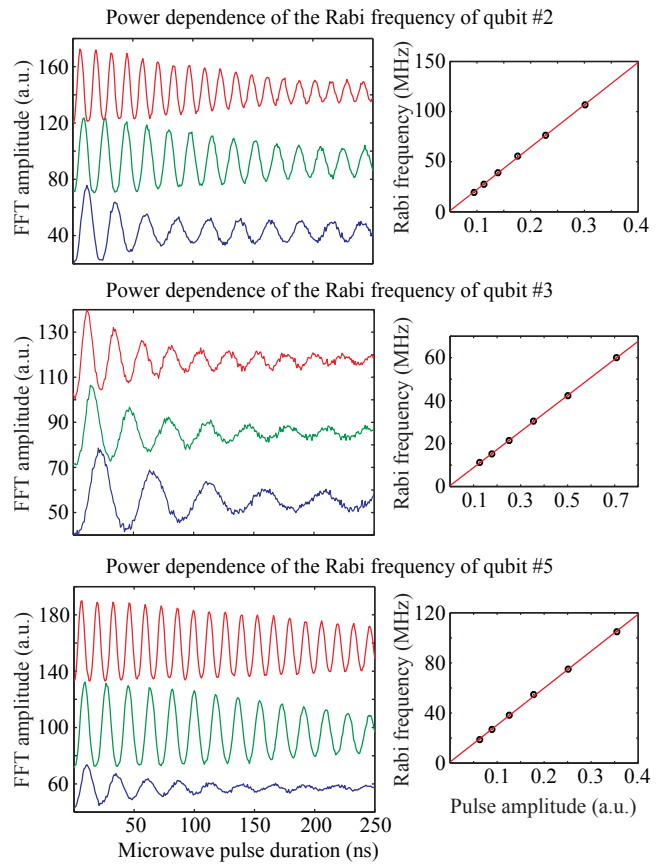}
\end{center}
\caption{
  (color online). 
  Simultaneous manipulation and detection of three qubits. 
  Left plots: 
    Rabi oscillations at several powers; 
    traces are vertically offset for better visibility; 
    curves with the same color/offset (blue-bottom, green-center, red-top) are measured simultaneously using the FDM technique described in the main text. 
  Right plots: 
    Rabi oscillation frequency versus power of the excitation tone; 
    the error bars are smaller than the size of the dots.
}
\label{fig:rabi}
\end{figure}

In the next set of experiments, we tuned the uniform flux coil and two compact local coils placed above the sample to bias three qubits at their symmetry points. 
Limiting the number of qubits to three was necessary because of the lack of additional (on-chip) coils and not due to the readout technique itself. 
After setup of the readout pulse to probe circuits number \#2, 3 and 5, we performed a spectroscopy of all three qubits simultaneously. 
A continuous microwave excitation signal of varying frequencies was applied to the sample and a pulsed three-tone probe signal was applied every $10\,\mathrm{\mu s}$.
When the excitation frequency matches the gap between the ground and first excited states of a qubit, the instantaneous dispersive shift of the center frequency of the corresponding resonator switches between positive and negative, thus changing the mean amplitude and phase of the transmitted probe tone. 
Figure \ref{fig:multi_spec} shows the spectra of three qubits measured in parallel. 

Finally, we performed simultaneous manipulation with time resolved measurements on three qubits. 
Here, we used individual microwave excitations for every qubit, which were added together via a power combiner. 
We note that the complete excitation chain could be replaced by a reference source and a mixer controlled by a single arbitrary waveform generator with sufficient bandwidth to drive all qubits, similar to FDM readout tone generation. 
Measurement data are reported in Fig.~\ref{fig:rabi}. 
All three qubits were simultaneously driven by individual excitation tones and the readout was performed in parallel using the FDM protocol.
Every qubit can be Rabi-driven at a different power. 
Left panels of Fig.~\ref{fig:rabi} present Rabi oscillations at three different powers for all qubits. 
The measured linear power dependences of Rabi oscillations reported on the right panels in Fig.~\ref{fig:rabi} are in excellent agreement with theory.

The reported technique can be scaled up to read hundreds of qubits on a chip. 
The minimum channel spacing is determined by the energy relaxation rates $\gamma$ of the qubits and $\kappa$ of the resonators.
$\kappa$ is dominated by the coupling to the feedline, and is a design parameter of the system.
Large values of $\kappa$ increase the signal-to-noise ratio of individual readouts\cite{Blais-PRB-04} but also enhance qubit decay due to spontaneous emission.
A channel spacing of $1.5\kappa$ is required to keep the crosstalk between adjacent readout channels below $-10\,\mathrm{dB}$, a spacing of $5\kappa$ reduces the crosstalk to $-20\,\mathrm{dB}$\cite{Jerger-EPL-2011}.
Energy relaxation of a qubit causes a frequency modulation of its readout resonator, which induces sidebands at multiples of $\Delta\omega_r$ from the bare resonator frequency.
A conservative estimate for the additional bandwidth required due to this modulation is\cite{Carson-PIRE-22} $2(\Delta\omega_r+2\gamma)$. 
Our current parameters allow 20 qubits per GHz of readout bandwidth with less than −20 dB of crosstalk and 60 qubits per GHz of readout bandwith with less than -10 dB of crosstalk. 
Furthermore, this number could be increased even more by lowering the coupling of the readout resonators to the feedline.

In conclusion, we developed a multiplexed readout scheme for superconducting qubits whose complexity does not depend on the number of qubits that are measured. 
In particular, we demonstrated parallel spectroscopy of six flux qubits as well as the independent manipulation and simultaneous readout of three of them. 
Despite its simplicity, this readout technique can be easily scaled up to measure registers consisting of hundreds of qubits. 

This work was supported by the EU project SOLID, the Deutsche Forschungsgemeinschaft (DFG) and the State of Baden-W\"urttemberg through the DFG-Center for Functional Nanostructures (CFN) within subproject B3.4.

\end{document}